\documentclass[superscriptaddress,twocolumn,nofootinbib,preprintnumbers,amsmath,amssymb,floatfix]{revtex4}

\usepackage{graphicx}
\usepackage{dcolumn}
\usepackage{bm}
\usepackage{xcolor}
\usepackage{subfigure}
\usepackage{enumitem}
\usepackage{ulem}

\begin{document}

\preprint{}

\title{Impact of resonance regeneration and decay on the net-proton fluctuations in a hadron resonance gas}

\author{Marlene Nahrgang}
\email{marlene.nahrgang@phy.duke.edu}
\affiliation{Department of Physics, Duke University, Durham, NC 27708-0305, USA}
\affiliation{Frankfurt Institute for Advanced Studies (FIAS), Ruth-Moufang-Str.~1, 60438 Frankfurt am Main, Germany}

\author{Marcus Bluhm}
\affiliation{Department of Physics, North Carolina State University, Raleigh, NC 27695, USA}

\author{Paolo Alba}
\affiliation{Dipartimento di Fisica, Universit\`{a} degli Studi di Torino \& INFN, Sezione di Torino, via Pietro Giuria 1, 10125 Torino, Italy}

\author{Rene Bellwied}
\affiliation{Department of Physics, University of Houston, Houston, TX 77204, USA}

\author{Claudia Ratti}
\affiliation{Department of Physics, University of Houston, Houston, TX 77204, USA}
 

\begin{abstract}
We investigate net-proton fluctuations as important observables measured in heavy-ion collisions within the hadron resonance gas (HRG) model. Special emphasis is given to effects which are a priori not inherent in a thermally and chemically equilibrated HRG approach. In particular, we point out the importance of taking into account the successive regeneration and decay of resonances after the chemical freeze-out, which lead to a randomization of the isospin of nucleons and thus to additional fluctuations in the net-proton number. We find good agreement between our model results and the recent STAR measurements of the higher-order 
moments of the net-proton distribution. 
\end{abstract}


\maketitle

\section{Introduction}

Relativistic heavy-ion collisions have contributed tremendously to our understanding of strongly interacting matter at high temperatures $T$ and net-baryon densities $n_B^{({\rm net})}$. 
The energy densities, which are locally reached in the  experiments~\cite{Arsene:2004fa,Back:2004je,Adams:2005dq,Adcox:2004mh,Chatrchyan:2012np,Abelev:2012rv}, are high enough to create a deconfined, strongly coupled plasma of quarks and gluons, in accordance with first-principle lattice QCD calculations~\cite{Cheng:2009zi,Borsanyi:2010cj}. 
The latter also showed that the transition from this deconfined phase to confined, hadronic matter is an analytic crossover for vanishing $n_B^{({\rm net})}$~\cite{Aoki:2006we}. 
In the confined phase, HRG model and lattice QCD results agree remarkably well with each other for the equilibrium thermodynamics~\cite{Karsch:2003vd,Karsch:2003zq}. Moreover, within statistical hadronization model analyses, experimental data on particle production are quite successfully described by corresponding thermal abundances calculated in HRG models with only a few freeze-out parameters for all collision energies ranging from AGS to the LHC, see e.g.~\cite{BraunMunzinger:2003zd,Cleymans:2005xv,Andronic:2011yq} and references therein. Thereby, the application of HRG models to characterize the bulk properties of hadronic matter is based on the assumption that after hadronization a thermally and chemically equilibrated system of strongly interacting hadrons is formed, which is well described in terms of a non-interacting gas of hadrons and resonances~\cite{Venugopalan:1992hy}. 

Recently, fluctuation observables have attracted much attention within theoretical and experimental studies. 
In fact, (higher-order) moments of particle multiplicity distributions provide an excellent opportunity to reveal more details of the collision process and, thus, of the phase structure of QCD. In particular, strictly conserved quantum numbers (charges) of the strong interaction, like baryon number ($B$), electric charge ($Q$) and strangeness ($S$) expressed in terms of their net-numbers $N_B^{({\rm net})}$, $N_Q^{({\rm net})}$ and $N_S^{({\rm net})}$, are of interest in heavy-ion collisions. 
For the conjectured critical point, for example, one expects a non-monotonic behavior in the fluctuations of net-baryon number and net-electric charge~\cite{Berges:1998rc,Halasz:1998qr,Stephanov:1998dy,Stephanov:1999zu,Hatta:2002sj}. 
Higher-order moments should be especially sensitive to critical phenomena if the correlation lengths grow in the vicinity of the critical point~\cite{Stephanov:2008qz}. However, non-equilibrium, dynamical effects such as critical slowing down can limit the growth of the correlation lengths~\cite{Berdnikov:1999ph} and thus influence the behavior of the higher-order moments~\cite{Kitazawa:2013bta}. For vanishing $n_B^{({\rm net})}$ fluctuation observables have also received revived attention because of the possibility to extract freeze-out parameters from first-principles by comparing experimental data to lattice QCD results~\cite{Karsch:2012wm,Bazavov:2012vg,Borsanyi:2013hza,Mukherjee:2013lsa}. 

In 2010 and 2011, the RHIC facility has engaged in the search for the critical point and the exploration of the QCD phase diagram by running a beam energy scan program with center-of-mass collision energies per nucleon-nucleon pair of $\sqrt{s}=7.7,\, 11.5,\, 19.6,\, 27,\, 39,\, 62.4$ and $200$~GeV. 
Recently, results on the net-proton fluctuations in terms of ratios of higher-order cumulants of the net-proton distribution were reported~\cite{Adamczyk:2013dal}. Due to the fact that isospin-fluctuations remain finite at the critical point, the critical fluctuations in the net-baryon number are directly imprinted in the net-proton fluctuations~\cite{Hatta:2003wn}. Sources of finite and non-equilibrium fluctuations can, however, significantly hide the critical fluctuations in the net-proton number as compared to net-baryon number fluctuations~\cite{Kitazawa:2011wh,Kitazawa:2012at}. 

Fluctuation observables have been investigated in various theoretical baseline studies within the HRG model~\cite{Begun:2006jf,Karsch:2010ck,Fu:2013gga,Garg:2013ata} or in transport approaches~\cite{Schuster:2009jv,Sahoo:2012wn}. 
In this work, we compare different HRG model calculations of net-proton fluctuations in a grandcanonical ensemble study by systematically including various refinements: a restriction from net-baryon number fluctuations to net-proton fluctuations, the application of experimentally realized kinematic cuts, an inclusion of the effects of strong resonance decays as well as of isospin-changing interactions of the nucleons with thermal pions after the chemical freeze-out. This study is of importance for future investigations of critical fluctuations produced in dynamical models of heavy-ion collisions in two ways: it serves as a non-critical baseline, and the considered refinements can also be applied to critical fluctuations.

We treat resonance decays as in~\cite{Begun:2006jf,Fu:2013gga}, but split the full contributions of the decaying resonances to the final net-proton fluctuations into two parts: an average and a probabilistic part. The average contributions stem from the thermal fluctuations in the numbers of resonances only, where fixed numbers of decay products are assumed, which are determined by the average branching ratios. The additional, probabilistic contributions account for the probabilistic character of the decay process implying fluctuations in the actual numbers of decay products. From the average contributions only, which we rederive via appropriate derivatives of the pressure, we observe a significant deviation of the net-proton fluctuations from the Skellam limit. Considering furthermore the probabilistic contributions from resonance decays as well as isospin-changing scatterings of nucleons with thermal pions via intermediate $\Delta$-resonances, which both lead to 
additional fluctuations in the net-proton number, is however important. 
A consistent treatment of these effects, especially of the latter via the Kitazawa-Asakawa (KA) formalism~\cite{Kitazawa:2011wh,Kitazawa:2012at}, can reconcile HRG model calculations for net-proton fluctuations with the experimental data on the same level as net-baryon number fluctuations calculated in a full HRG model. 

Besides the effects considered explicitly in this work, further possible sources of fluctuations exist, which are important when comparing to experimental data:
1) in heavy-ion collisions the global net-baryon number, net-electric charge and net-strangeness are conserved exactly and not only on average as in a grandcanonical ensemble. This can cause large effects on the fluctuations~\cite{Schuster:2009jv,Bzdak:2012an}. In fact, it is only due to the limitations in the kinematic acceptance that one can assume the measured data to be describable within a grandcanonical ensemble. In studies using the UrQMD transport model, which accounts for the micro-canonical nature of the individual scatterings, it was shown that while net-baryon number fluctuations are strongly affected, the net-proton fluctuations are affected only at lower $\sqrt{s}$~\cite{Schuster:2009jv}, which is in agreement with the latest UrQMD calculations performed by the STAR collaboration~\cite{Adamczyk:2013dal}. 
2) Experimental reconstruction efficiencies and impurities also lead to fluctuations. The STAR net-proton data in~\cite{Adamczyk:2013dal} is corrected for reconstruction efficiencies, except for the ratio of second- to first-order cumulants. Our studies showed that the difference between the uncorrected and the corrected results for this ratio is negligible as long as the reconstruction efficiencies for protons and anti-protons exceed 70\%, which is the case for all $\sqrt{s}$. For the corrections in the higher-order cumulants a binomial distribution was assumed in~\cite{Adamczyk:2013dal}. The purity of the proton sample is 98\%. In line with~\cite{Ono:2013rma}, one can estimate that the remaining 2\%, assuming they are Poisson-distributed, affect the results for the ratios of the third- to second-order cumulants by 1\% and of the fourth- to second-order cumulants only by 0.1\%. 

This paper is organized as follows: in the next section we discuss aspects of the HRG model, which is used throughout this work. Section~\ref{sec:flucHRG} presents step-by-step our results for the ratios of the higher-order cumulants of the net-proton distribution including kinematic cuts, resonance decays and isospin-changing reactions. Conclusions follow in section~\ref{sec:conclusions}. 

\section{Hadron resonance gas}\label{sec:HRG}

We perform our study of net-proton fluctuations within a HRG model, which includes $113$ mesons, $103$ baryons and their corresponding anti-baryons up to masses of approximately $2$~GeV, as used in~\cite{Huovinen:2009yb} for the construction of a QCD equation of state. 
The equilibrium pressure $P$ is given by the sum of the partial pressures of all particle species $i$ included in the model 
\begin{equation}
 P/T^4=\frac{1}{VT^3}\sum_{i}\ln{\cal Z}_{m_i}^{M/B}(V,T,\mu_B,\mu_Q,\mu_S)\, ,
\label{eq:pressureHRG}
\end{equation}
where the natural logarithms of the grandcanonical partition functions ${\cal Z}_{m_i}^{M/B}$ for mesons (M, upper signs) and (anti-)baryons (B, lower signs) are given by a momentum integral, 
\begin{equation}
 \ln{\cal Z}_{m_i}^{M/B}=\mp\frac{Vd_i}{(2\pi)^3}\int {\rm d}^3k\, \ln(1\mp z_i\exp(-\epsilon_i/T))\, .
\label{eq:lnZHRG}
\end{equation}
Here, the single-particle energy reads $\epsilon_i=\sqrt{k^2+m_i^2}$ with the particle mass $m_i$, $d_i$ is the degeneracy factor, $V$ is the volume and 
\begin{equation}
 z_i=\exp((B_i\mu_B+Q_i\mu_Q+S_i\mu_S)/T)\equiv\exp(\mu_i/T)
\label{eq:fug}
\end{equation}
is the fugacity. In Eq.~(\ref{eq:fug}), the $\mu_X$ denote the chemical potentials conjugate to the net-densities of the conserved charges $X$ and $X_i=B_i,\, Q_i,\, S_i$ represent the quantum numbers of baryon charge, electric charge and strangeness of each particle species. The partial derivative of the pressure with respect to the particle chemical potential $\mu_i$ gives the density of particles $i$, 
\begin{equation}
 n_i(T,\mu_i)=\frac{d_i}{(2\pi)^3}\int {\rm d}^3k\, f_{\rm FD/BE}(T,\mu_i)
\label{eq:particledensity}
\end{equation}
with the Fermi-Dirac (FD) or Bose-Einstein (BE) distribution function $f_{\rm FD/BE}$ for (anti-)baryons or mesons. 
Summing $n_i$ multiplied by $X_i$ over all particle species $i$, one obtains the net-density of the conserved charge $X$, $n_X^{({\rm net})}=\sum_iX_i\,n_i\equiv\sum_iX_i\langle N_i\rangle/V$, which corresponds to $\left.\partial P/\partial \mu_X\right|_T$. 

In a HRG model, particles are usually considered as pointlike, a point of view which we also take. The influence of repulsive van-der-Waals forces on the fluctuations, included in the model in form of excluded volumes of the particles, has been discussed in \cite{Fu:2013gga}. 

The chemical composition of a HRG in local thermal and chemical equilibrium is then determined by the independent chemical potentials $\mu_i$ of each individual species, their masses and the temperature. Due to the rapid expansion of the created matter, however, the density decreases, which leads to an enhancement of the particle mean free path. At a given set of thermodynamic parameters ($T^{\rm fo}, \mu_B^{\rm fo},\mu_Q^{\rm fo},\mu_S^{\rm fo}$), reactions like baryon--anti-baryon annihilation (e.g. $p\bar{p}\to\pi\pi\pi\pi\pi$) or pion production (e.g. $N\pi\to N^*(1520)\to\Delta\pi\to N\pi\pi$ and $\pi\pi\to\omega\pi\to\pi\pi\pi\pi$) and their corresponding back-reactions become too rare to maintain chemical equilibrium among different particle species. 
This set of parameters describes the chemical freeze-out, an instant at which chemical equilibrium is lost, the chemical composition of the gas is frozen-out and after which only elastic scatterings occur frequently enough to maintain local thermal equilibrium until even these become too rare and the particles start to stream freely after the kinetic freeze-out. 

A more realistic picture of the hadronic stage assumes that chemical equilibrium is not completely lost just after the chemical freeze-out~\cite{Bebie:1991ij}: as long as $T$ is high enough, specific reactions in the form of resonance regenerations and decays, (e.g.~$\pi\pi\to\rho\to\pi\pi$, $K\pi\to K^*\to K\pi$ and $p\pi\to\Delta\to p\pi$), continue to occur at a significant rate. Resonances are consequently still in chemical equilibrium with their decay products. 
However, the final numbers $\hat{N}_h$ (i.e.~primordial numbers as present at the chemical freeze-out plus resonance decay contributions) of those hadron species $h$, which do not decay strongly within the duration of the hadronic stage, are conserved because the aforementioned particle number changing reactions are inefficient after the chemical freeze-out. 
The hadronic matter is, thus, in a state of partial chemical equilibrium. Consequently, the chemical potentials of all stable hadrons, $\mu_h$, become $T$-dependent, while the chemical potentials of the resonances, $\mu_R$, become functions of the $\mu_h$ via $\mu_R=\sum_h\mu_h\langle n_h\rangle_R$. Here, the sum runs over all stable hadrons and $\langle n_h\rangle_R\equiv\sum_r b_r^R\,n_{h,r}^R$ is the decay-channel averaged number of hadrons $h$ produced in the decay of resonance $R$, where $b_r^R$ is the branching ratio of the decay-channel $r$ of $R$ and $n_{h,r}^R=0,1,...$ is the number of $h$ formed in that specific decay-channel. With decreasing $T$, eventually all resonances decay either directly or via a decay-chain into stable hadrons and are not regenerated anymore. 

In this work, the chemical freeze-out parameters are taken as an input. 
According to~\cite{Cleymans:2005xv}, the temperature is described by a polynomial function of $\mu_B$ via 
\begin{equation}
 T^{\rm fo}(\mu_B^{\rm fo})=a-b\,(\mu_B^{\rm fo})^2-c\,(\mu_B^{\rm fo})^4
\end{equation}
with $a=(0.166\pm0.002)$~GeV, $b=(0.139\pm0.016)$~GeV$^{-1}$ and $c=(0.053\pm0.021)$~GeV$^{-3}$. The baryon-chemical potential itself is given as a function of $\sqrt{s}$ in the form 
\begin{equation}
 \mu_B^{\rm fo}(\sqrt{s}\,)=\frac{d_B}{1+e_B\sqrt{s}}\, ,
\label{eq:muBfo}
\end{equation}
with $d_B=(1.308\pm0.028)$~GeV and $e_B=(0.273\pm0.008)$~GeV$^{-1}$. The $\sqrt{s}$-dependence of the electric charge and strangeness chemical potentials, $\mu_Q$ and $\mu_S$, has to be determined from requiring~\cite{Karsch:2010ck} 
\begin{align}
\label{eq:conditions1}
 n_S^{({\rm net})}(T,\mu_B,\mu_Q,\mu_S)&=0 \,,\\
\label{eq:conditions2}
 n_Q^{({\rm net})}(T,\mu_B,\mu_Q,\mu_S)&=x\, n_B^{({\rm net})}(T,\mu_B,\mu_Q,\mu_S)\, .
\end{align}
These conditions reflect the situation in a heavy-ion collision, namely the net-strangeness neutrality and the ratio of protons to baryons $x\simeq 0.4$ for Au+Au and Pb+Pb collisions present in the initial state. The equality for $n_Q^{({\rm net})}$ in Eq.~(\ref{eq:conditions2}) takes also into account that due to the lack of stopping at high beam energies the interesting mid-rapidity region is almost isospin symmetric. This is ensured through the $\sqrt{s}$-dependence of $\mu_B$ and the correspondingly small $n_B^{\rm (net)}$ at high $\sqrt{s}$. In the same form as in Eq.~(\ref{eq:muBfo}), $\mu_Q^{\rm fo}$ and $\mu_S^{\rm fo}$ can be approximated parametrically as functions of $\sqrt{s}$. The parameters in our HRG model approach are $d_Q=-0.0202$~GeV, $e_Q=0.125$~GeV$^{-1}$ and $d_S=0.224$~GeV, $e_S=0.184$~GeV$^{-1}$. 

\begin{figure}
 \includegraphics[width=0.44\textwidth]{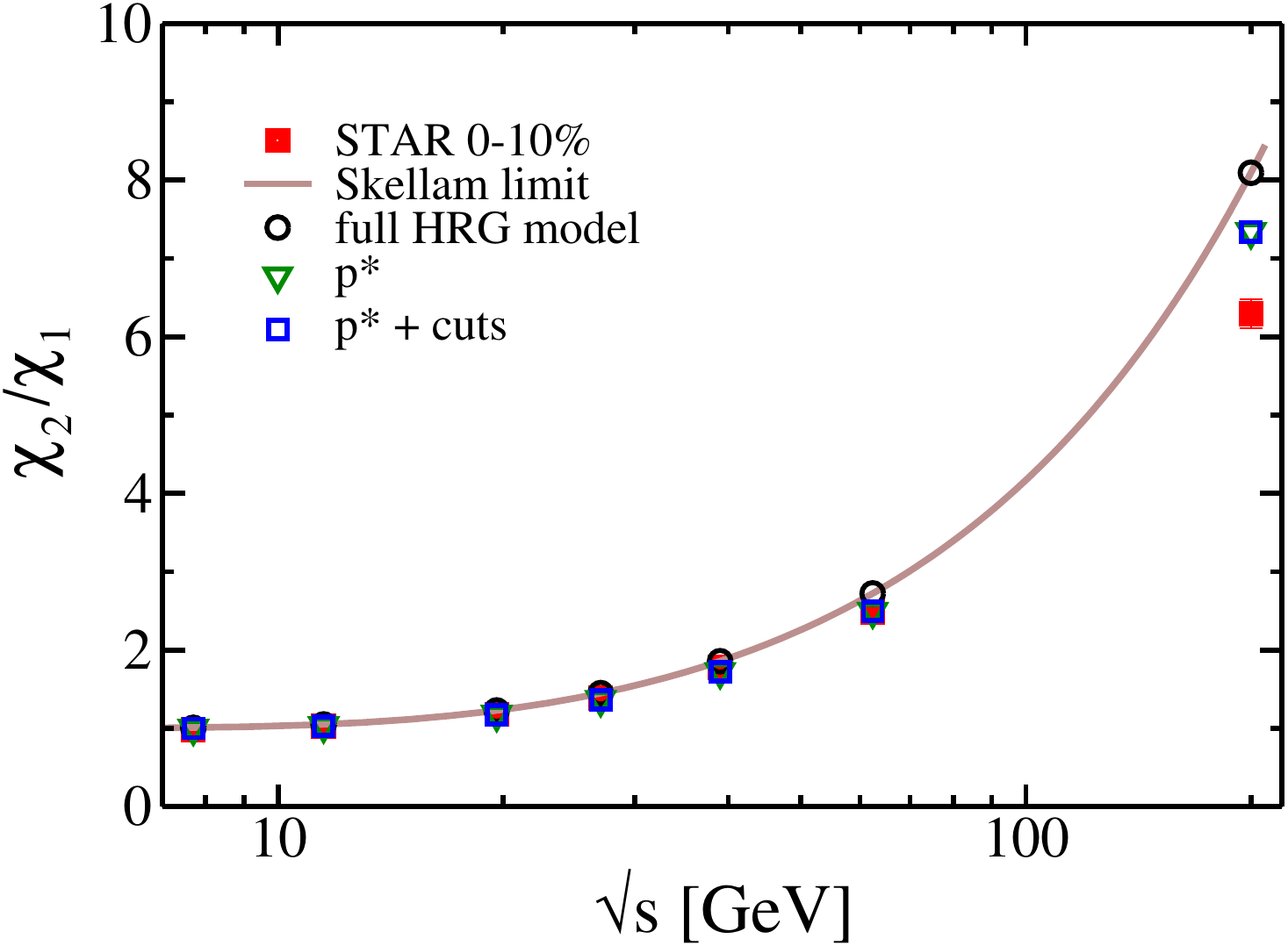}\\
 \includegraphics[width=0.44\textwidth]{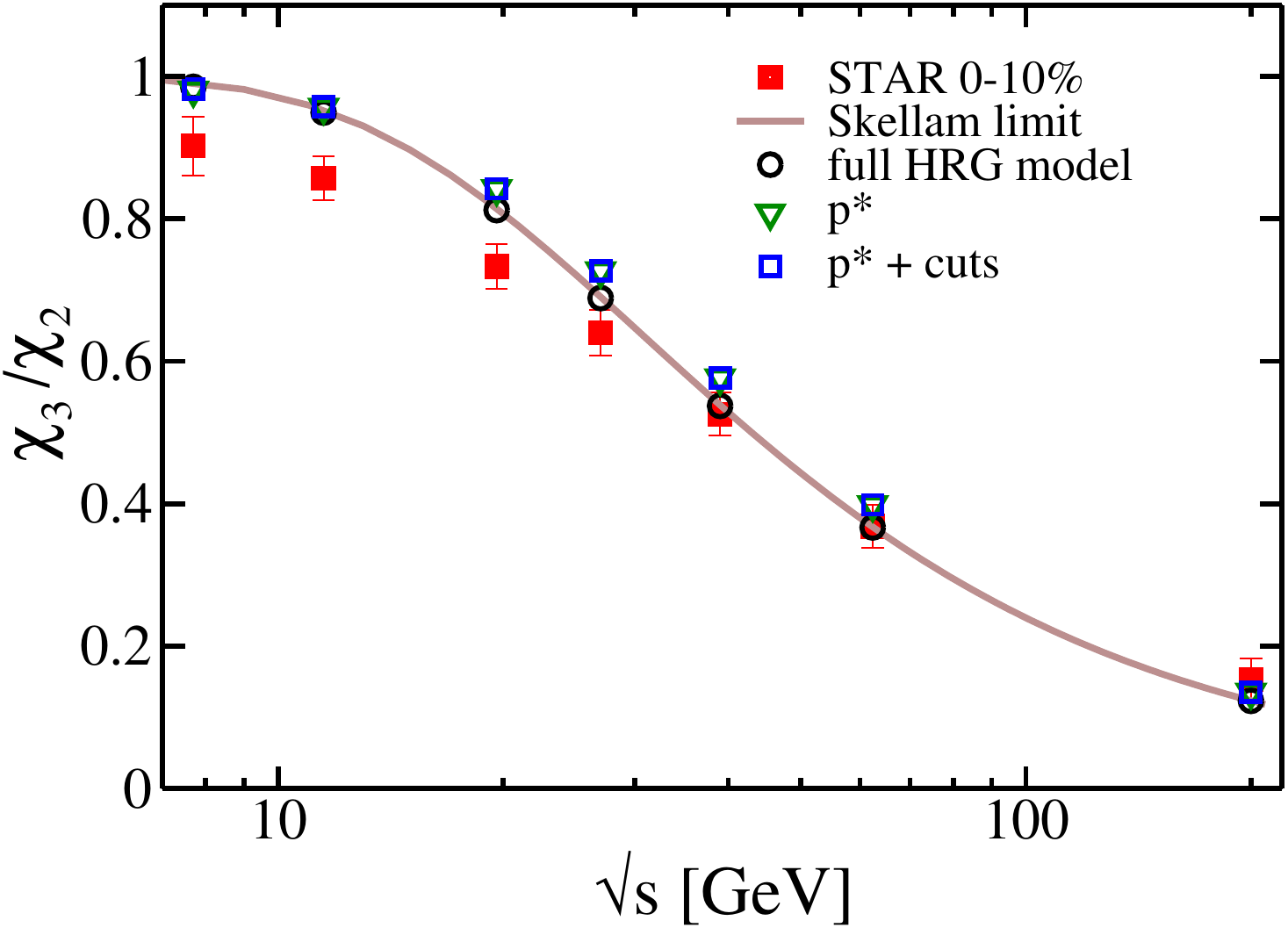}\\
 \includegraphics[width=0.44\textwidth]{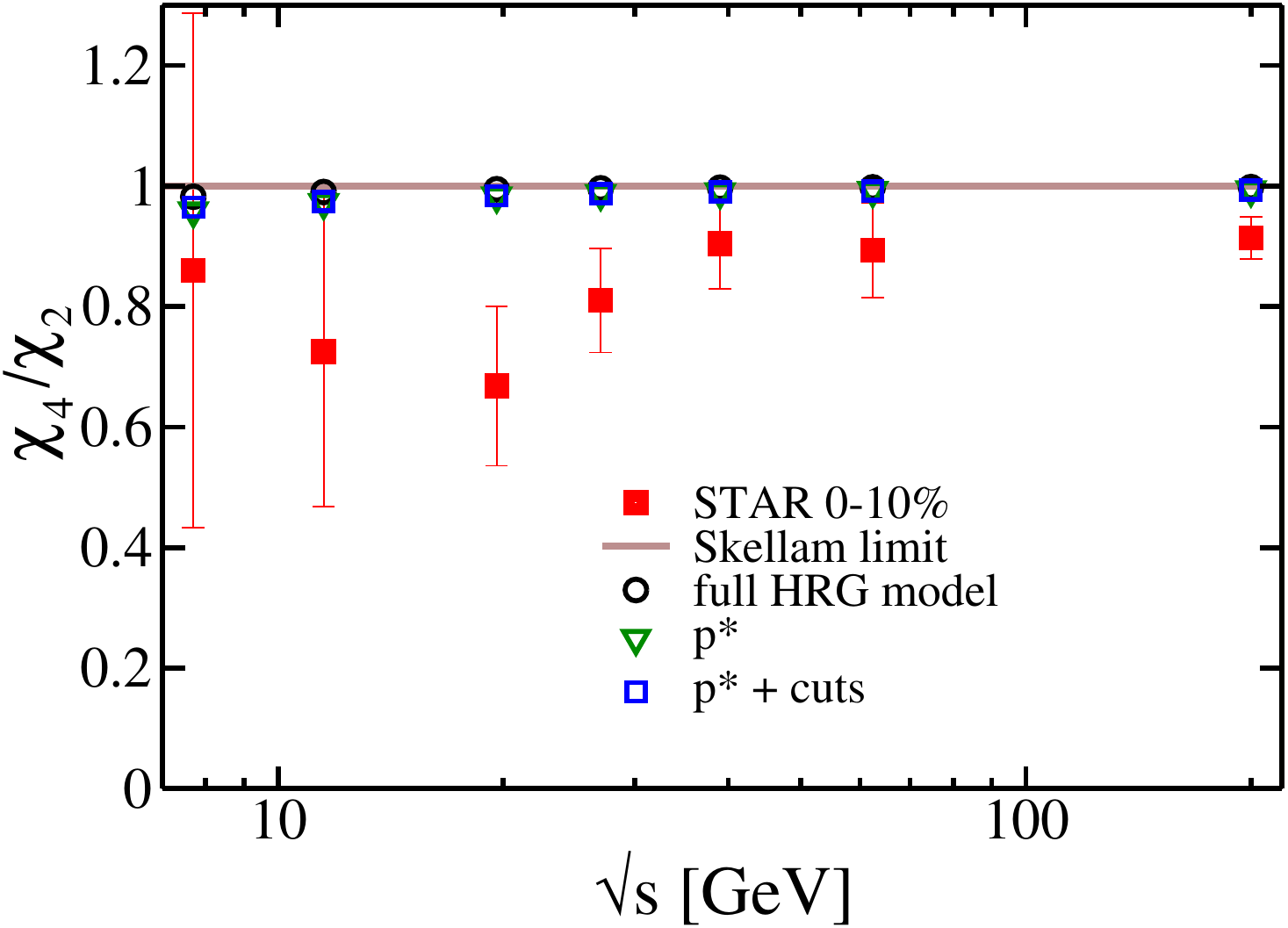}
 \caption{Beam energy dependence of the susceptibility ratios $\chi_2/\chi_1$ (left panel), $\chi_3/\chi_2$ (middle panel) and $\chi_4/\chi_2$ (right panel) which are connected to the experimental observables as in Eq.~(\ref{eq:ratiocumulants}). The full squares depict experimental data on net-proton fluctuations as measured by the STAR collaboration~\cite{Adamczyk:2013dal} for the two most central collision classes ($0$-$10$\%). These are compared to specific HRG model results: empty circles show the susceptibility ratios for the net-baryon number fluctuations in our full HRG model containing 103 baryon species and their anti-baryons. The empty triangles show the corresponding ratios for the net-proton fluctuations considering primordial protons and anti-protons only, while the empty squares highlight the additional, negligibly small influence of applying kinematic cuts as explained in the text. The solid curves show the corresponding Skellam limits for a Boltzmann gas of baryons and anti-baryons.}
 \label{fig:fig1}
\end{figure}

\section{Fluctuations in a hadron resonance gas}\label{sec:flucHRG}
 
The susceptibilities describing fluctuations in the number of particles of species $i$ in a thermally and chemically equilibrated HRG are defined by derivates of the scaled pressure in Eq.~(\ref{eq:pressureHRG}) with respect to the corresponding particle chemical potential 
\begin{equation}
 \chi_l^{(i)}=\left.\frac{\partial^l(P/T^4)}{\partial(\mu_i/T)^l}\right|_{T}
\label{eq:chi_n}
\end{equation}
and can be related to the cumulants of the distribution of that particle species via 
\begin{align}
 \chi_1^{(i)}&=\frac{1}{VT^3}\langle N_i\rangle_c=\frac{1}{VT^3} \langle N_i\rangle\, ,\\
 \chi_2^{(i)}&=\frac{1}{VT^3}\langle(\Delta N_i)^2\rangle_c=\frac{1}{VT^3} \langle (\Delta N_i)^2\rangle\, ,\\
 \chi_3^{(i)}&=\frac{1}{VT^3}\langle(\Delta N_i)^3\rangle_c=\frac{1}{VT^3} \langle (\Delta N_i)^3\rangle\, ,\\
 \chi_4^{(i)}&=\frac{1}{VT^3}\langle(\Delta N_i)^4\rangle_c \nonumber \\
 &\equiv\frac{1}{VT^3} \left(\langle (\Delta N_i)^4\rangle-3\langle (\Delta N_i)^2\rangle^2\right)\, ,
\end{align}
where the first three cumulants are equal to the corresponding central moments, while the fourth cumulant is given by a combination of fourth and second central moments, and $\Delta N_i=N_i-\langle N_i\rangle$. 

For an equilibrium HRG model in the grandcanonical ensemble formulation, thermally produced and non-interacting particles and anti-particles are uncorrelated. Thus, the susceptibilities of the net-distributions can be expressed via the susceptibilities of particle and anti-particle distributions as 
\begin{equation}
 \chi_l^{({\rm net},i)}= \chi_l^{(i)}+(-1)^l\,\chi_l^{(\bar{i})}\, .\\
\label{eq:independprod}
\end{equation} 
Particular ratios of the susceptibilities can be expressed in terms of the mean $M=\langle N\rangle$, the variance $\sigma^2=\langle(\Delta N)^2\rangle$, the skewness $S= \langle (\Delta N)^3\rangle/\langle (\Delta N)^2\rangle^{3/2}$ and the kurtosis $\kappa=\langle (\Delta N)^4\rangle/\langle (\Delta N)^2\rangle^2-3$, for example 
\begin{equation}
 \frac{\chi_2}{\chi_1}=\frac{\sigma^2}{M},\quad \frac{\chi_3}{\chi_2}=S\sigma,\quad \frac{\chi_4}{\chi_2}=\kappa\sigma^2\, .
\label{eq:ratiocumulants}
\end{equation}
In these ratios, the experimentally unkown volume term cancels on average as well as the dependence on the particle numbers due to the central limit theorem. In general, volume fluctuations due to fluctuations in the initial collision geometry can influence the cumulant ratios \cite{Skokov:2012ds}. 

In Fig.~\ref{fig:fig1}, we compare the ratios in Eq.~(\ref{eq:ratiocumulants}) evaluated for different degrees of refinements of the HRG model to the  STAR data for central ($0$-$10$\%) collisions~\cite{Adamczyk:2013dal}. For the full HRG model, we calculate the susceptibility ratios for the net-baryon number, to which all $103$ baryons included in the model and their anti-baryons contribute. One finds a rather good agreement with the experimental data with some deviations around the dip in $\kappa\sigma^2$, for the lower beam energies in $S\sigma$ and for the $\sqrt{s}=200$~GeV data point in $\sigma^2/M$. Our results also agree nicely with the ones obtained in~\cite{Karsch:2010ck} and with the same quantities calculated within the Boltzmann approximation to the HRG model, which gives a good description for $(m_i-\mu_B)/T\gg1$. In this approximation the primordial net-baryon and net-proton distributions are given by Skellam distributions, as indicated for net-baryon number by the solid 
curves in Fig.~\ref{fig:fig1}. 
If one restricts the particle sample to primordial protons and anti-protons, the results remain similar to those for the net-baryon number fluctuations in the full HRG model. Small differences are seen only for high $\sqrt{s}$ in $\sigma^2/M$, for intermediate $\sqrt{s}$ in $S\sigma$ and for low $\sqrt{s}$ in $\kappa\sigma^2$. 

\subsection{Experimental cuts}\label{sec:expcuts}

The experimental phase-space coverage is limited in rapidity $y$ and transverse momentum $k_T$ according to the detector design and the demands from reconstruction efficiency and particle identification. 
In their recent analysis~\cite{Adamczyk:2013dal}, the STAR collaboration considered the following kinematic acceptance cuts:  $|y|\leq0.5$ and $0.4\, {\rm GeV}\leq k_T\leq 0.8$~GeV with full azimuthal, i.e.~$\phi=2\pi$, coverage. 
For a meaningful comparison with the experimental data one should, therefore, aim at including these cuts in the model calculations, too. 

In~\cite{Garg:2013ata}, it was proposed to model acceptance cuts by limiting the integration range in Eqs.~(\ref{eq:lnZHRG}) and~(\ref{eq:particledensity}) accordingly. For this, the momentum variables $(k_x,k_y,k_z)$ are transformed into $(k_T,y,\phi)$, which implies replacing the integration measure ${\rm d}^3k$ by $k_T\sqrt{k_T^2+m_i^2}\cosh(y)\,{\rm d}k_T\,{\rm d}y\,{\rm d}\phi$ and the single-particle energies $\epsilon_i$ by $\cosh(y)\sqrt{k_T^2+m_i^2}$. 
By applying the same strategy for primordial protons and anti-protons, we obtain results for the net-proton fluctuations as shown in Fig.~\ref{fig:fig1}. In the Boltzmann approximation to the HRG model one can separate the fugacity factors from the momentum integrals. One thus expects that the influence of cuts in $y$ and $k_T$ is negligible, since in this approximation the cutted integrals cancel in the considered cumulant ratios of the net-proton distribution. This observation is indeed made in Fig.~\ref{fig:fig1} and remains true for all following refinements of the HRG model, which we investigate in this paper. We note here that 
while in our model approach all momentum integrals are evaluated at the chemical freeze-out, the final kinematics, which is subject to the acceptance cuts, is determined at the lower, kinetic freeze-out temperature. 
In principle, a study of the evolution of the thermal distributions of the particles until the kinetic freeze-out, taking elastic scatterings in the thermally equilibrated hadronic phase into account, would be needed in order to implement the kinematic cuts more realistically. 

\subsection{Resonance decays}

Resonances play an important role in the evolution of the created strongly interacting, hadronic matter and their decays can significantly influence the final numbers of the stable hadrons as well as fluctuations therein. Just after the chemical freeze-out, when the matter is in a state of partial chemical equilibrium, the $\mu_R$ depend on the $\mu_h$. This dependence provides a means to derive the average influence of the resonance decays on the fluctuations in the final particle numbers: 
considering the derivative of $P/T^4$ with respect to $\mu_h/T$ as in Eq.~(\ref{eq:chi_n}), but keeping in mind that only the chemical potentials $\mu_h$ are independent of each other, while the $\mu_R$ depend on $\mu_h$, one arrives at 
\begin{equation}
 VT^3\left.\frac{\partial(P/T^4)}{\partial(\mu_h/T)}\right|_{T}= \langle N_h\rangle + \sum_R \langle N_R\rangle \langle n_h\rangle_R\, .
\label{eq:chi1Final}
\end{equation}
This is equivalent to the mean of the final number $\langle \hat{N}_h\rangle$ of the stable hadron species $h$ after resonance decays discussed in section~\ref{sec:HRG}~\cite{Bebie:1991ij}. 
In Eq.~(\ref{eq:chi1Final}), $\langle N_h\rangle$ and $\langle N_R\rangle$ denote the means of the primordial numbers of hadrons and resonances and the sum runs over all the resonances in the model. In agreement with the QCD equations of state constructed in~\cite{Bluhm:2013yga}, we consider here $26$ different particle species as stable, namely $\pi^0$, $\pi^+$, $\pi^-$, $K^+$, $K^-$, $K^0$, $\overline{K}^0$, $\eta$ as well as $p$, $n$, $\Lambda^0$, $\Sigma^+$, $\Sigma^0$, $\Sigma^-$, $\Xi^0$, $\Xi^-$, $\Omega^-$ and their anti-baryons. This implies that contributions stemming from weak decays are not taken into account, which is in accordance with the experimental analysis~\cite{Adamczyk:2013dal}. 

In the following, we concentrate on the fluctuations in the final numbers of protons and anti-protons only. 
Making use of the $\mu_p$-dependence of the $\mu_R$, the cumulants of the final distribution of protons (the same expressions hold for anti-protons when $p$ is replaced by $\overline{p}$) follow from derivatives of $P/T^4$ with respect to $\mu_p/T$ and read 
\begin{align}
\label{eq:Chi1Netp}
 \langle \hat{N}_{p}\rangle&=\langle N_{p}\rangle+ \sum_R \langle N_R\rangle \langle n_{p}\rangle_R\, ,\\
\label{eq:Chi2Netp}
 \langle (\Delta\hat{N}_{p})^2\rangle&=\langle (\Delta N_{p})^2\rangle+ \sum_R \langle (\Delta N_R)^2\rangle \langle n_{p}\rangle_R^2\, ,\\
\label{eq:Chi3Netp}
 \langle (\Delta\hat{N}_{p})^3\rangle&=\langle (\Delta N_{p})^3\rangle+ \sum_R \langle (\Delta N_R)^3\rangle \langle n_{p}\rangle_R^3\, ,\\
\label{eq:Chi4Netp}
 \langle (\Delta\hat{N}_{p})^4\rangle_c&=\langle (\Delta N_{p})^4\rangle_c+ \sum_R \langle (\Delta N_R)^4\rangle_c \langle n_{p}\rangle_R^4\, .
\end{align}
The related susceptibilities are given by 
\begin{align}
\label{eq:ChiNetpAdd}
 \hat{\chi}_l^{(p)}=\chi_l^{(p)}+\sum_R \chi_l^{(R)}\langle n_{p}\rangle_R^l\, .
\end{align}
These expressions account for the contributions arising from the thermal fluctuations in the numbers of primordial resonances if one assumes fixed, average numbers of produced protons as determined by the branching ratios of the resonance decays. 

\begin{figure}
 \includegraphics[width=0.44\textwidth]{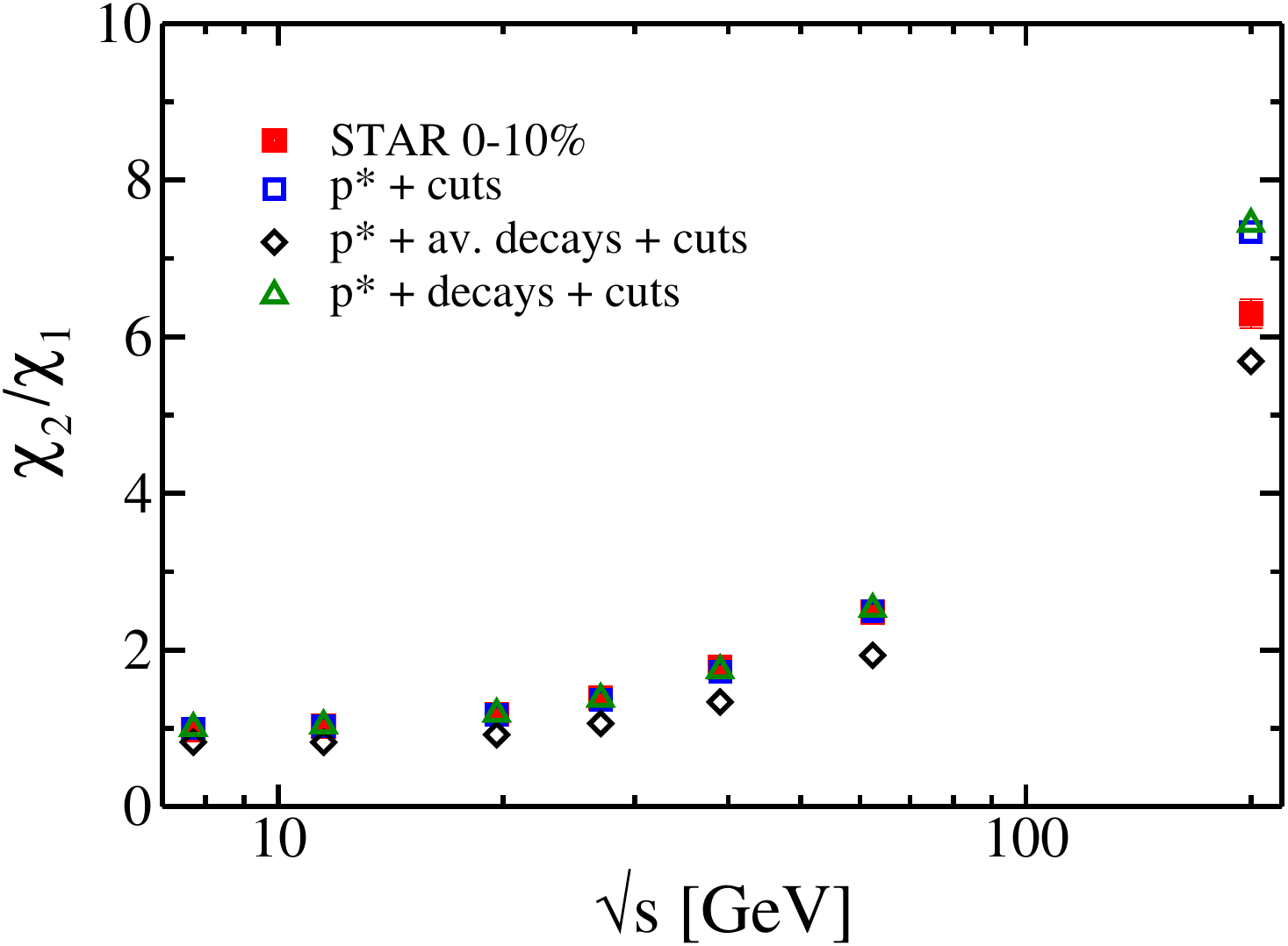}\\
 \includegraphics[width=0.44\textwidth]{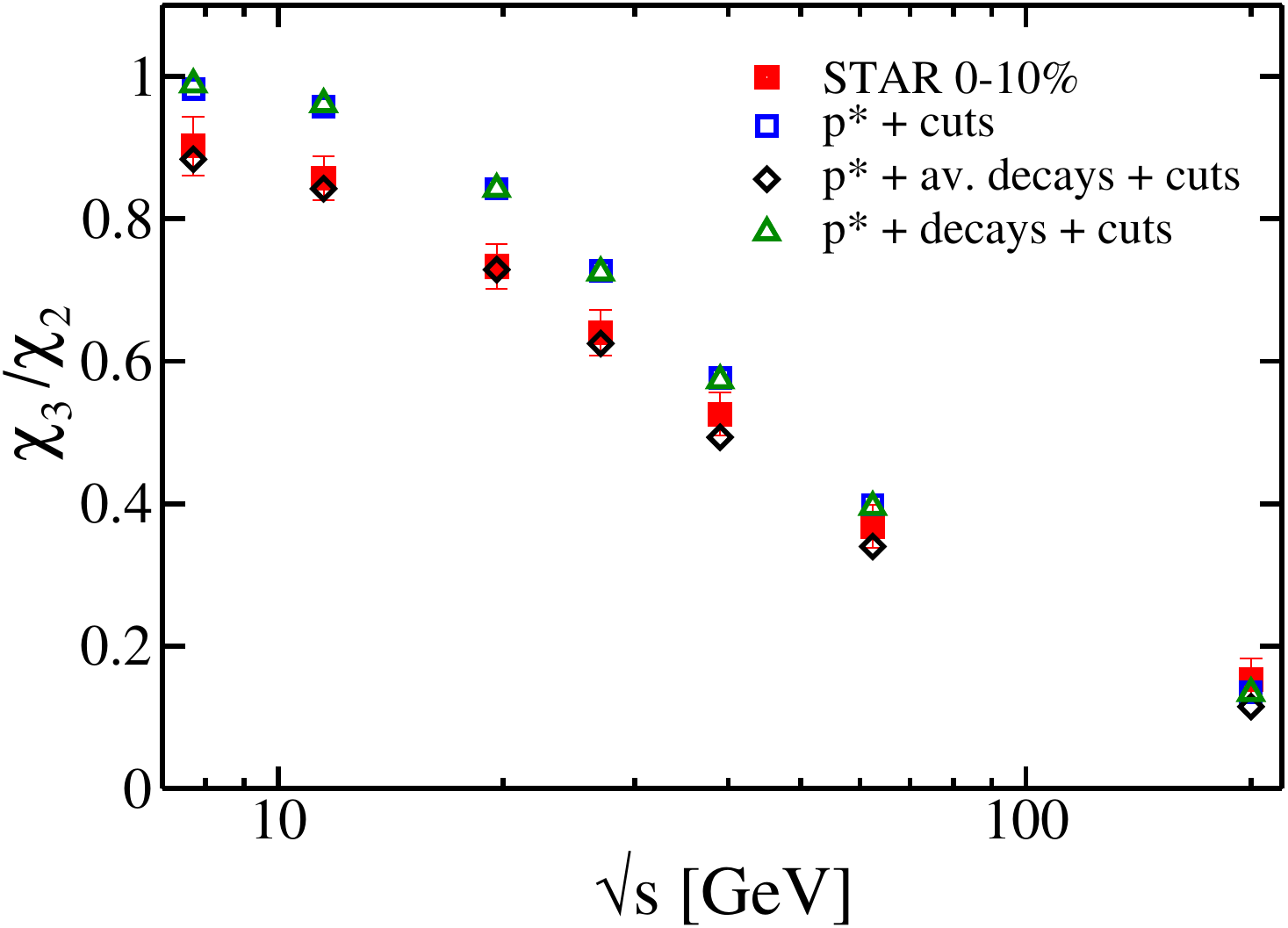}\\ 
 \includegraphics[width=0.44\textwidth]{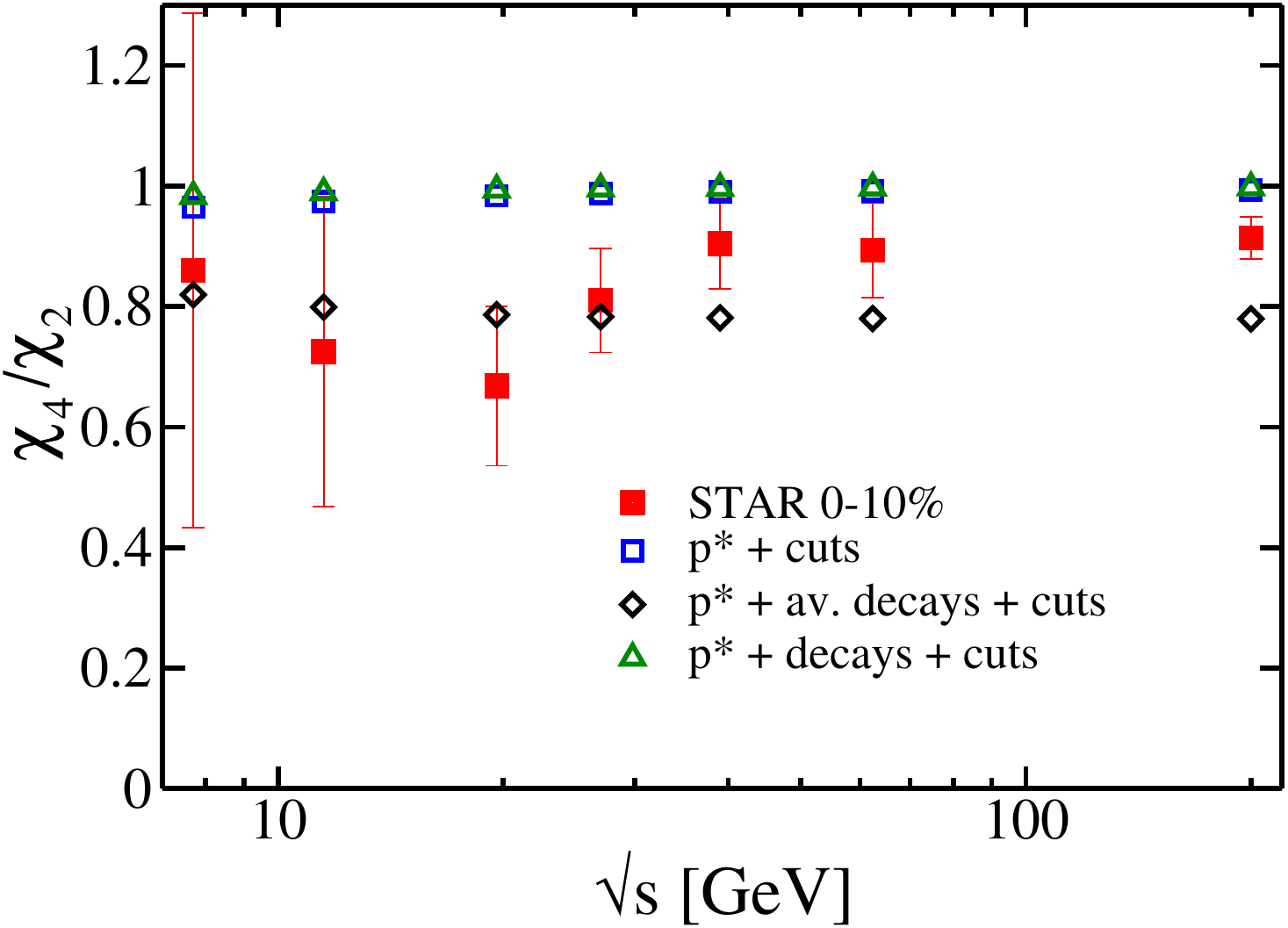}
 \caption{Similar to Fig.~\ref{fig:fig1}, but considering additional refinements of the HRG model calculations: the empty squares show the same as in Fig.~\ref{fig:fig1}, while the empty diamonds highlight the average influence of the resonance decays on the net-proton fluctuations. The empty triangles show the full impact of resonance decays, including the probabilistic contributions.}
 \label{fig:fig2}
\end{figure}

Resonance decays are, however, probabilistic processes. For example, the decay of the resonance $\Delta^+(1232)$ yields only on average $\langle n_p\rangle_{\Delta^+}$ protons, $\langle n_n\rangle_{\Delta^+}$ neutrons, $\langle n_{\pi^+}\rangle_{\Delta^+}$ positive pions and $\langle n_{\pi^0}\rangle_{\Delta^+}$ neutral pions, where we use $\langle n_p\rangle_{\Delta^+}=0.669$, $\langle n_n\rangle_{\Delta^+}=0.331$, $\langle n_{\pi^+}\rangle_{\Delta^+}=0.331$ and $\langle n_{\pi^0}\rangle_{\Delta^+}=0.663$. 
In reality, the actual numbers of decay products follow a multinomial distribution, which itself results in fluctuations in the final particle numbers. In order to take this into account one is required to go beyond thermal derivatives. The full impact of resonance decays was studied for the first two cumulants in~\cite{Begun:2006jf} and for the third and fourth cumulant in~\cite{Fu:2013gga} starting from the general probability distribution for the decay of resonances. For a grandcanonical ensemble, the corresponding cumulants of the final proton distribution read 
\begin{align}
\label{eq:Chi1NetpZWB}
 \langle \hat{N}_{p}\rangle&=\langle N_{p}\rangle + \sum_R \langle N_R\rangle \langle n_{p}\rangle_R\, ,\\
\nonumber
 \langle (\Delta\hat{N}_{p})^2\rangle&=\langle (\Delta N_{p})^2\rangle + \sum_R \langle (\Delta N_R)^2\rangle \langle n_{p}\rangle_R^2 \\
\label{eq:Chi2NetpZWB}
 &\,\,\,\,\,\, + \sum_R \langle N_R\rangle \langle (\Delta n_{p})^2\rangle_R \, ,\\
\nonumber
 \langle (\Delta\hat{N}_{p})^3\rangle&=\langle (\Delta N_{p})^3\rangle + \sum_R \langle (\Delta N_R)^3\rangle \langle n_{p}\rangle_R^3 \\
\nonumber
 &\,\,\,\,\,\, + 3\sum_R \langle (\Delta N_R)^2\rangle \langle n_p\rangle_R \langle (\Delta n_{p})^2 \rangle_R \\
\label{eq:Chi3NetpZWB}
 &\,\,\,\,\,\, + \sum_R \langle N_R\rangle \langle (\Delta n_{p})^3\rangle_R\, ,\\
\nonumber
 \langle (\Delta\hat{N}_{p})^4\rangle_c&=\langle (\Delta N_{p})^4\rangle_c + \sum_R \langle (\Delta N_R)^4\rangle_c \langle n_{p}\rangle_R^4 \\
\nonumber
 &\,\,\,\,\,\, + 6\sum_R \langle (\Delta N_R)^3\rangle \langle n_p\rangle_R^2 \langle (\Delta n_{p})^2 \rangle_R \\
\nonumber
 &\,\,\,\,\,\, + \sum_R \langle (\Delta N_R)^2\rangle \bigg[3\,\langle (\Delta n_{p})^2 \rangle_R^2 \\ 
\nonumber
 &\,\,\,\,\,\,\,\,\,\,\,\,\,\,\, + 4\,\langle n_p\rangle_R \langle (\Delta n_{p})^3 \rangle_R \bigg] \\
 &\,\,\,\,\,\, + \sum_R \langle N_R\rangle \langle (\Delta n_{p})^4\rangle_{R,c} \, .
\label{eq:Chi4NetpZWB}
\end{align}
In general, the factors $\langle (\Delta n_h)^2\rangle_R$,  $\langle (\Delta n_h)^3\rangle_R$ and  $\langle (\Delta n_h)^4\rangle_{R,c}$ vanish exactly for those resonances, which have only one decay-channel, or for which the number of formed hadrons $n_{h,r}^R$ of species $h$ is the same in each decay-channel $r$. For protons this is the case for $\Delta^{++}(1232)$ and $\Delta^{++}(1930)$, which only have one decay-channel, and for all mesonic resonances because they do not decay into protons. 
Equations~(\ref{eq:Chi1NetpZWB})-(\ref{eq:Chi4NetpZWB}) clearly contain the average fluctuation contributions from resonance decays as derived above in Eqs.~(\ref{eq:Chi1Netp})-(\ref{eq:Chi4Netp}). 

Two remarks are in order here. First, since in our framework primordial protons and anti-protons are uncorrelated, and no baryonic (anti-baryonic) resonance decays into an anti-proton (proton), the formula of independent production in Eq.~(\ref{eq:independprod}) remains valid for the susceptibilities of the net-proton distribution even when resonance decays are included. Second, we apply the same kinematic cuts to the resonances as to the primordial protons and anti-protons although experimentally the decay products are subject to the kinematic acceptance cuts. 
In general, the kinematics is different for the decay products and for the resonances. A Monte-Carlo study in~\cite{Jeon:1999gr} showed, however, that for cuts in rapidity this difference has only a negligible influence of less than $1\%$ on the results. In addition, due to the elastic scatterings in the thermally equilibrated hadronic phase it seems to be more likely that the kinematic cuts affect the primordial (anti-)protons in the same manner as the (anti-)protons stemming from resonance decays. 

In Fig.~\ref{fig:fig2}, the influence of resonance decays on the net-proton fluctuations is exhibited and contrasted with our results for primordial protons and anti-protons without resonance decay contributions as shown in Fig.~\ref{fig:fig1}. The average contributions of the resonance decays result in large deviations from our results for the net-baryon number fluctuations in the full HRG model (up to 20\% in $\sigma^2/M$, 10\% in $S\sigma$ and 20\% in $\kappa\sigma^2$, cf.~Fig.~\ref{fig:fig1}). This is a consequence of the fact that for most of the proton-producing resonances $0<\langle n_p\rangle_R<1$, such that the resonance decay contributions in Eqs.~(\ref{eq:Chi1Netp})-(\ref{eq:ChiNetpAdd}) induce significant differences from Poissonian behavior in the final proton (and equivalently anti-proton) susceptibilities (most easily seen in the right panel of Fig.~\ref{fig:fig2}, where with the average resonance decay contributions $\kappa\sigma^2<1$). 
Comparing to the experimental data one arrives at slightly different conclusions for the different susceptibility ratios: while the agreement with the data for $\sigma^2/M$ is globally worsened, the description of $S\sigma$ at lower beam energies improves. For $\kappa\sigma^2$ the deviations from the Skellam limit are clearly seen. The agreement with the data is slightly improved at lower $\sqrt{s}$, where the error bars are large, and worse at higher $\sqrt{s}$. 

The additional, probabilistic contributions balance the effect of the average resonance decay contributions and the final net-proton fluctuations with the full impact of resonance decays come close to the original results for the primordial net-proton fluctuations. This is a consequence of the fact that for each resonance $R$ the actual number $n_{p,r}^R$ of produced protons for a given decay-channel $r$ is either $0$ or $1$ (similarly for anti-protons) such that $\langle n_p^l\rangle_R\equiv \sum_r b_r^R(n_{p,r}^R)^l=\langle n_p\rangle_R$. We stress that this situation is notably different for pions. Within the Boltzmann approximation, the full resonance decay contributions to each cumulant in the Eqs.~(\ref{eq:Chi2NetpZWB})-(\ref{eq:Chi4NetpZWB}) individually add then up to $\sum_R \langle N_R\rangle \langle n_{p}\rangle_R$, as in Eq.~(\ref{eq:Chi1NetpZWB}), 
such that the final proton (or anti-proton) number follows the Poisson distribution.

\subsection{Isospin-randomization}

In addition to the resonance decays further important interactions take place after the chemical freeze-out. 
Notably, processes of the form 
\begin{align}
\label{eq:delta2}
 p(n)+\pi^0(\pi^+)&\to\Delta^{+}\to n(p)+\pi^+(\pi^0)\, ,\\
\label{eq:delta3}
 p(n)+\pi^-(\pi^0)&\to\Delta^{0}\to n(p)+\pi^0(\pi^-) 
\end{align}
via an intermediate $\Delta$-resonance (preferably $\Delta(1232)$) can change the isospin-identity of the nucleons in the hadronic phase. Similar processes occur for the anti-nucleons. These reactions do not alter average quantities and are, thus, irrelevant for statistical hadronization model fits to the ratios of particle yields, but they certainly affect the higher-order fluctuations. Due to this additional source of stochastic fluctuations one can expect that any original distribution of (anti-)protons will be pushed closer toward the Poisson limit.

The importance of the above final state interactions for relating the measured 
net-proton fluctuations to the theoretically more interesting net-baryon number fluctuations has first been realized in~\cite{Kitazawa:2011wh,Kitazawa:2012at}. The probability that after one cycle of the processes in Eqs.~(\ref{eq:delta2}) and~(\ref{eq:delta3}) a nucleon (anti-nucleon) has changed its isospin-identity is $4/9$. After two cycles this probability is approximately $50\%$. 
Thus, the (anti-)nucleon isospin gets completely randomized if one can assume that the (anti-)nucleons undergo at least two of these cycles in the hadronic phase between chemical and kinetic freeze-out. 
The number of protons among a given number of nucleons is then given by a binomial distribution and the probability distributions for the numbers of protons, anti-protons, neutrons and anti-neutrons factorize. 

\begin{figure}
 \includegraphics[width=0.44\textwidth]{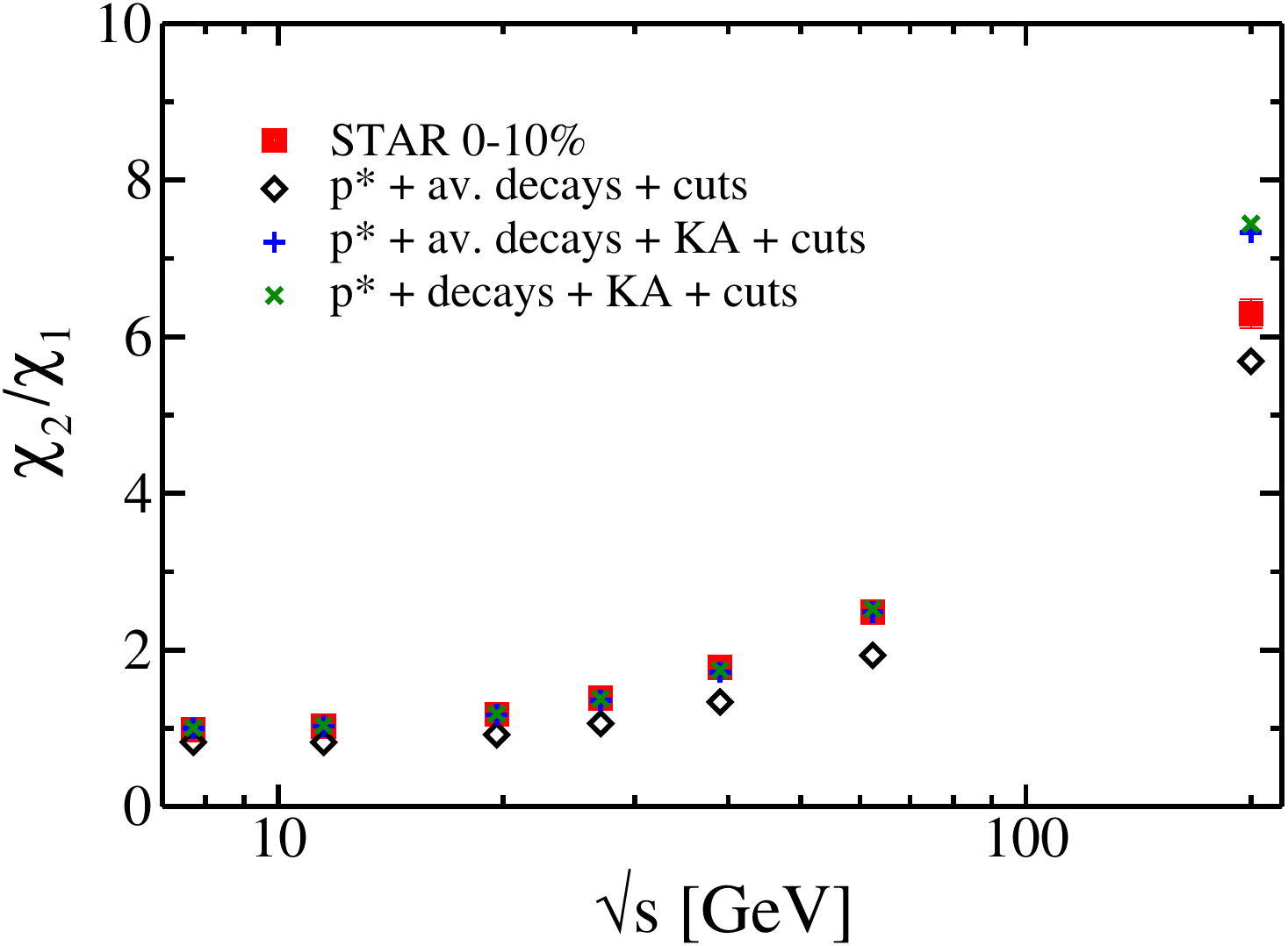}\\
 \vspace{3mm}

 \includegraphics[width=0.44\textwidth]{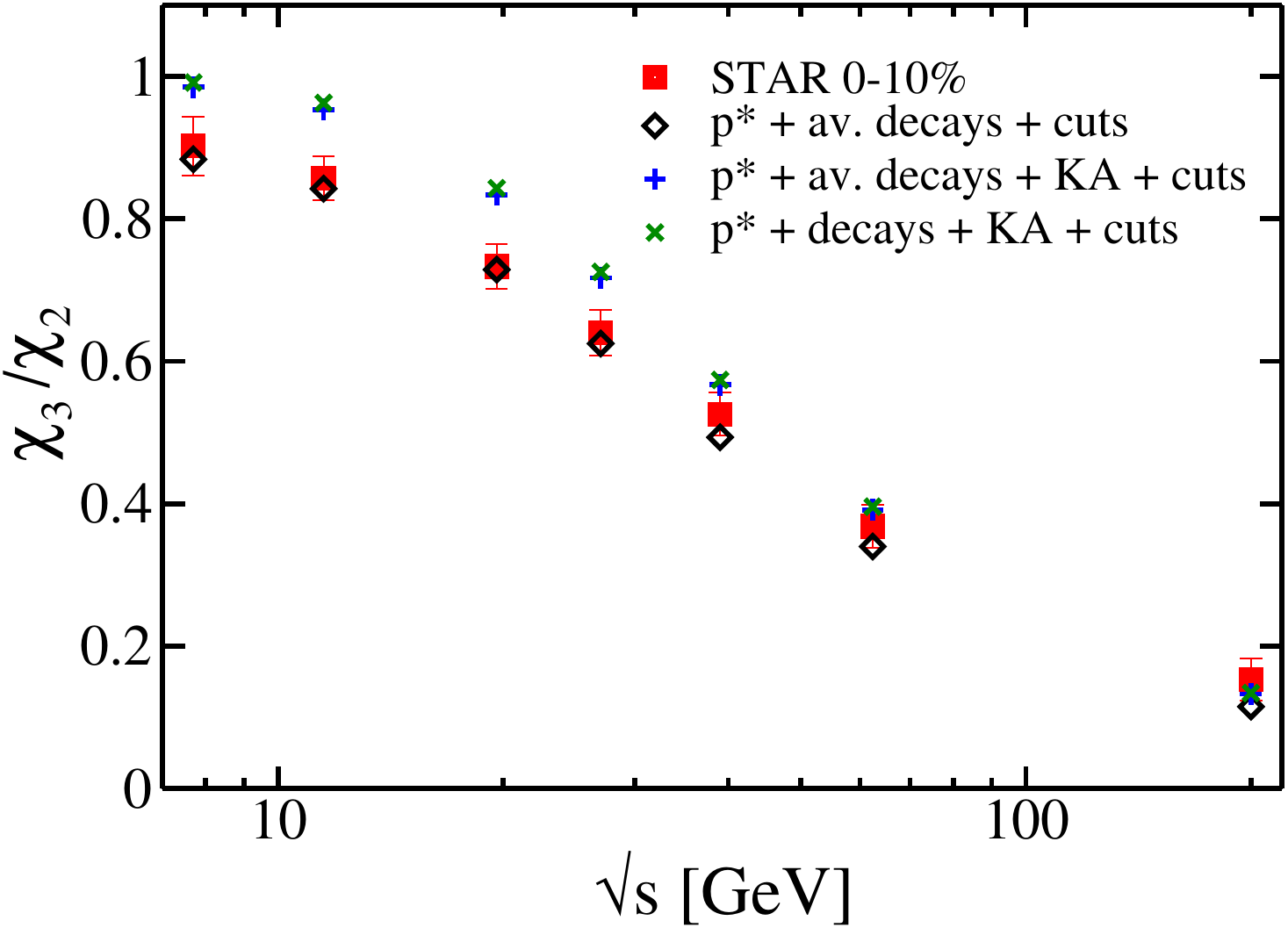}\\
 \vspace{3mm}

 \includegraphics[width=0.44\textwidth]{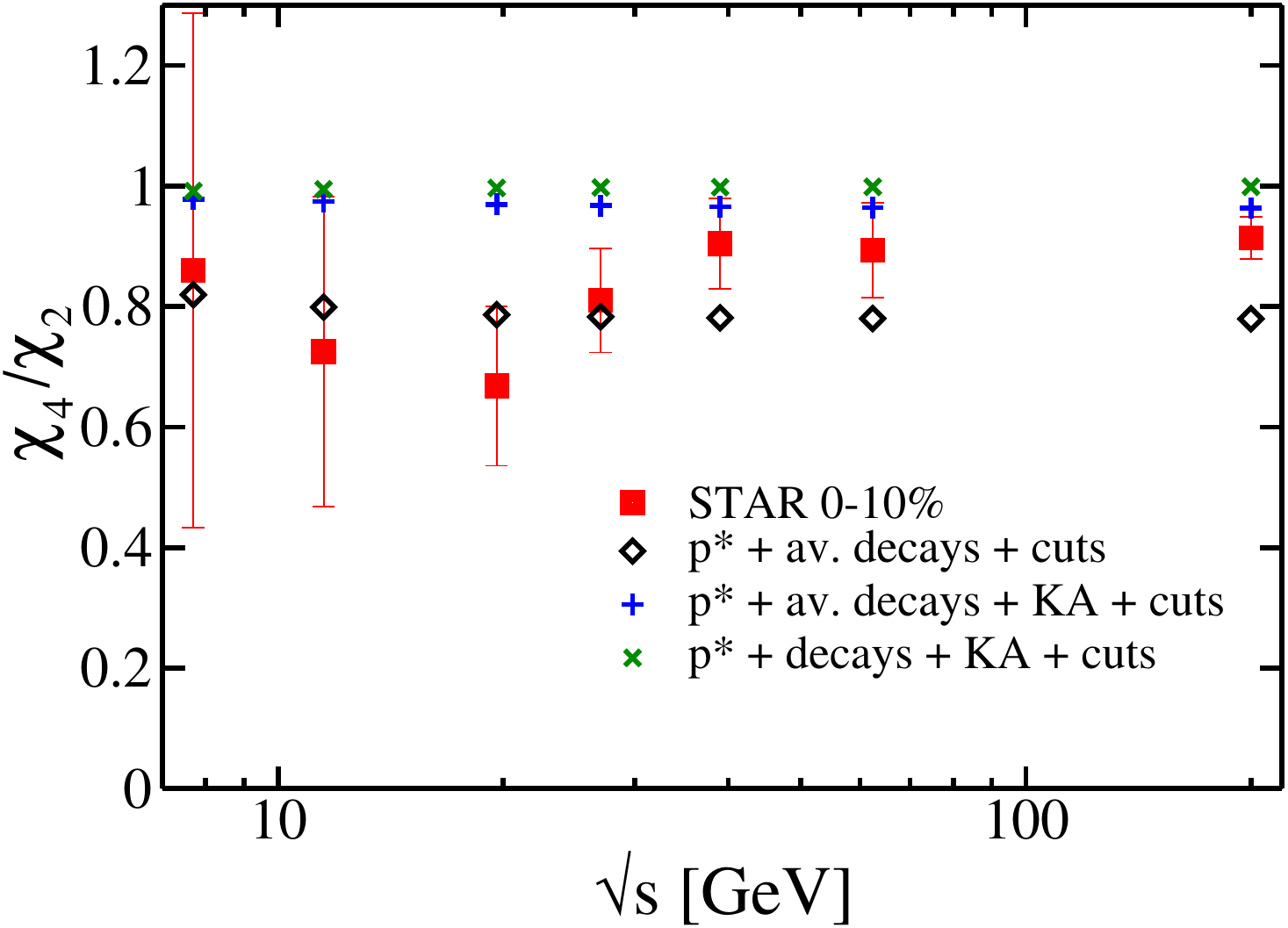}
 \caption{Similar to Fig.~\ref{fig:fig1}, but comparing three different HRG model calculations for net-proton fluctuations with each other: the empty diamonds depict the same results as in Fig.~\ref{fig:fig2}, while the plus-signs show the additional impact of the isospin-randomization described by the Kitazawa-Asakawa (KA) formalism. Results for the KA-formalism with full resonance decay contributions as input are shown as crosses.}
 \label{fig:fig3}
\end{figure}

In order to be efficient, the isospin-randomization requires short mean times for the processes in Eqs.~(\ref{eq:delta2}) and~(\ref{eq:delta3}) compared to the duration of the stage between chemical and kinetic freeze-out. While the regeneration time for the $\Delta(1232)$-resonances depends crucially on the pion density and is approximately $3-4$~fm in the temperature range between $150-170$~MeV~\cite{Kitazawa:2011wh,Kitazawa:2012at}, their lifetimes are about $1.8$~fm. 
According to the transport calculations in~\cite{Nonaka:2006yn}, the pion density and the duration of the hadronic phase are sufficient to fully randomize the isospin of the nucleons for $\sqrt{s}\gtrsim10$~GeV, cf.~\cite{Kitazawa:2011wh,Kitazawa:2012at}. At top-RHIC energy, STAR measurements~\cite{Adams:2004yc,Adams:2006yu} suggest, however, that the system expands very fast and that the duration of the hadronic stage is only of about $4-6$~fm. Full isospin-randomization might, thus, no longer be achieved at top-RHIC and LHC energies. We do not discuss this question further but present our results as an upper limit for the impact of the isospin randomization at all beam energies considered in this work.

While the main purpose of the work in~\cite{Kitazawa:2011wh,Kitazawa:2012at} was to obtain the net-baryon number fluctuations from the net-proton fluctuations, here, we apply the equations~(36)-(40) from Ref.~\cite{Kitazawa:2012at} in order to reconstruct the net-proton fluctuations. 
In these equations, we use the cumulants of the nucleon and anti-nucleon distributions instead of the cumulants of the baryon and anti-baryon distributions because weak decays are excluded in our approach. For the final nucleon number $\hat{N}_N=\hat{N}_p+\hat{N}_n$, including the average contributions from resonance decays, the related susceptibilities follow from 
\begin{equation}
 \hat{\chi}_l^{(N)}=\chi_l^{(p)}+\chi_l^{(n)}+\sum_R\chi_l^{(R)}\left(\langle n_p\rangle_R+\langle n_n\rangle_R\right)^l \,.
\label{eq:chinucl}
\end{equation}
For $l>1$, the cumulants of the final nucleon distribution are, thus, not a simple sum of the final proton and neutron cumulants. In fact, for all non-strange baryonic resonances the sum $\langle n_p\rangle_R+\langle n_n\rangle_R=1$, such that the final nucleon cumulants are essentially given by the sum of the primordial proton, neutron and proton- and/or neutron-producing resonance cumulants. 

In Eq.~(\ref{eq:chinucl}), the probabilistic decay contributions are not included, since they are suppressed when we consider protons and neutrons (similarly anti-protons and anti-neutrons) together. In fact, only excited strange baryons, e.g.~$\Lambda(1520)$, which decay for example into $\Sigma\pi$, $\Lambda(\Sigma)\pi\pi$ or $p(n)K$ in different decay-channels, would contribute to the probabilistic part. 
In the effective description of complete isospin randomization the probabilistic decay of the $\Delta$-resonances is already included and it can be assumed that the additional effect of the probabilistic decay of further resonances as input to the KA-modifications is small.

In Fig.~\ref{fig:fig3}, we show our results which include the isospin-randomization via $\Delta$-resonance regeneration and decay. This effect is implemented independently of the beam energy. We observe a substantial improvement in the agreement with the data for $\sigma^2/M$ in comparison with the previous scenario (i.e.~primordial net protons plus average decay contributions and cuts), although for $\sqrt{s}=200$~GeV our result is above the measured value. This deviation is, however, smaller than the one for the net-baryon number fluctuations in the full HRG model (cf.~left panel of Fig.~\ref{fig:fig1}). The description of the experimental data for $S\sigma$ at lower $\sqrt{s}$ is less good than in the previous scenario, but comparable with the full HRG model (cf.~middle panel of Fig.~\ref{fig:fig1}). For higher $\sqrt{s}$ the experimental data for $\kappa\sigma^2$ is described slightly better than in the previous scenario. Overall, the 
global agreement with the data is slightly improved compared to the net-baryon number fluctuations in the full HRG model (cf.~Fig.~\ref{fig:fig1}) due to the improved description of $\sigma^2/M$. We have checked that the probabilistic decay contributions to the nucleon susceptibilities as input for the KA-modifications has only a small additional effect as seen by comparing the plus-signs and the crosses in Fig.~\ref{fig:fig3}.
We note, again, that the Kitazawa-Asakawa (KA) formalism is limited to $\sqrt{s}\gtrsim 10$~GeV and should, most likely, not be applied to the lowest beam energy of $\sqrt{s}=7.7$~GeV included in this study. 

\section{Conclusions}\label{sec:conclusions}

In this paper, we investigated systematically the influence of various refinements in the HRG model calculation of net-proton fluctuations and compared our results to the recent STAR data in~\cite{Adamczyk:2013dal}. Starting from the net-baryon number fluctuations in our full HRG model containing $103$ baryon species and their anti-baryons, we restricted the sample to primordial protons and anti-protons and determined the corresponding net-proton fluctuations. For the considered freeze-out parameters, these results agree well with the experimental data and are close to the Skellam limit. 

Unlike in studies of the conserved charges of QCD, resonance decays can become important for restricted particle samples.
We find that the average contributions from resonance decays are derivable within the framework of a HRG in partial chemical equilibrium. They induce significant deviations from Poissonian behavior in the (anti-)proton susceptibilities and worsen the agreement with the data for $\sigma^2/M$. On the contrary, the probabilistic character of the decay process, which cannot be accounted for by thermal derivatives, restores the results toward the Poisson limit by adding an additional source for fluctuations. A limitation of the momentum integrals in accordance with the kinematic acceptance cuts does not lead to visible changes of our results. 

Finally, we applied the Kitazawa-Asakawa formalism~\cite{Kitazawa:2011wh,Kitazawa:2012at} in order to reconstruct the net-proton susceptibilities from the nucleon and anti-nucleon susceptibilities, for which we either took only the average or the full resonance decay contributions into account. In this way, the effect of the isospin-randomization of \mbox{(anti-)nucleons} via intermediate $\Delta$-resonance regeneration and decay on the final net-proton fluctuations was analyzed quantitatively for the first time. 
We find that the agreement with the experimental data for $\sigma^2/M$ is mostly improved compared to the situation where only the average influence of the resonance decays is considered. The non-Poissonian signature in $\kappa\sigma^2$ induced by the average resonance decay contributions is obviously smoothed out through additional stochastic components such as resonance regeneration effects. 
We note that these results represent an upper limit obtained under the assumption of full isospin-randomization in the hadronic phase for all $\sqrt{s}$. The application of the KA-formalism was straightforward as all ingredients are directly calculable within the HRG model. By including all of these refinements we did not change a basic feature of the HRG model, namely the possibility of describing fluctuation observables with only two parameters, the freeze-out temperature, $T^{\rm fo}$, and the freeze-out baryon-chemical potential, $\mu_B^{\rm fo}$. 

We note that other non-critical effects on fluctuation results like volume fluctuations, efficiency corrections, excluded volume corrections in a HRG model and the global baryon-number conservation have the tendency to be more important for the ratios of higher-order susceptibilities and for lower $\sqrt{s}$. The separate impact of resoncance decay and regeneration, in contrast, shows up already in the lowest-order ratio, $\chi_2/\chi_1$, as discussed in this work. 

Our reconstructed results for the net-proton fluctuations in Fig.~\ref{fig:fig3} agree also very nicely for most of the beam energies with all three susceptibility ratios of the net-baryon number fluctuations in Fig.~\ref{fig:fig1}. The good agreement of the latter calculated in a full HRG model with the net-proton fluctuations measured by the STAR collaboration can, thus, be understood as a combined impact of resonance decays and isospin-randomization in the hadronic phase after the chemical freeze-out. 
This may also be seen as a reinforcement that any contribution going beyond a fully equilibrated hadron resonance gas could be washed out in the net-proton fluctuations as pointed out in~\cite{Kitazawa:2011wh,Kitazawa:2012at}. 

In future work it will be interesting to investigate how resonance decays and isospin randomization affect fluctuation signals from a potential phase transition, in particular the QCD critical point.

\section*{Acknowledgements}
M.~Nahrgang thanks the Yukawa Institute for Theoretical Physics, Kyoto University, where some fruitful discussions have motivated part of this work.
This work is supported by the Hessian LOEWE initiative Helmholtz International Center for FAIR, the Italian Ministry of Education, Universities and Research under the Firb Research Grant RBFR0814TT and the US Department of Energy grants DE-FG02-07ER41521 and DE-FG02-03ER41260. M.~Nahrgang acknowledges support by a fellowship within the Postdoc-Program of the German Academic Exchange Service (DAAD).

\end{document}